\documentclass[preprint,12pt]{article}

\usepackage{amsmath, amssymb,amsfonts,amsthm}

\usepackage{graphicx}

\usepackage{array}
\makeatletter
\newcommand{\thickhline}{%
    \noalign {\ifnum 0=`}\fi \hrule height 1pt
    \futurelet \reserved@a \@xhline}
\makeatother
\usepackage{tabularx}
\usepackage{longtable}
\usepackage[table]{xcolor}
\usepackage{multirow}

\usepackage{lscape}

\usepackage{float}

\usepackage{color}

\usepackage[a4paper, total={6in, 8in}]{geometry}

\usepackage{authblk}

\date{}
\begin{document}


\title{A machine learning procedure\\ to detect network attacks}

\author[1]{Davide Coppes}
\author[2]{Paolo Cermelli}
\affil[1]{Department of Physics, University of Torino, Italy}
\affil[2]{Department of Mathematics, University of Torino, Italy}


\maketitle
\begin{abstract}
The goal of this note is to assess whether simple machine learning algorithms can be used to determine whether and how a given network has been attacked. The procedure is based on the $k$-Nearest Neighbor and the Random Forest classification schemes, using both intact and attacked Erd\H{o}s-R\'enyi, Barabasi-Albert and Watts-Strogatz networks to train the algorithm. The types of attacks we consider here are random failures and maximum-degree or maximum-betweenness node deletion. Each network is characterized by a list of 4 metrics, namely the normalized reciprocal maximum degree, the global clustering coefficient, the normalized average path length and the assortativity: a statistical analysis shows that this list of graph metrics is indeed significantly different in intact or damaged networks. We test the procedure by choosing both artificial and real networks, performing the attacks and applying the classification algorithms to the resulting graphs: the procedure discussed here turns out to be able to distinguish between intact networks and those attacked by the maximum-degree of maximum-betweenness deletions, but cannot detect random failures. Our results suggest that this approach may provide a basis for the analysis and detection of network attacks.

\end{abstract}

\section{Introduction}

The relevance of the study of attacks to networks was already recognized at the early stages of network science, given their importance as a substantial threat to the  spread of information and the network integrity. The first studies in this field were able to identify a substantial difference in the response to attacks in dependence of the structure of the attacked network \cite{albert1,barabasibook, bellingeri1}: in fact, networks with exponential degree distribution, such as the Erd\H{o}s-R\'enyi graph, are equally sensitive to random failures and targeted attacks, while scale-free networks are almost insensitive to random failures, being severely disrupted by targeted attacks. Random failures are generally understood to consist in random deletion of nodes \cite{albert1, jalili1, norrenbrock1, almeira1, crucitti1}, while targeted attacks are usually performed by removing nodes according to some centrality measure, such as degree centrality \cite{albert1, bellingeri1, jalili1, norrenbrock1, almeira1, crucitti1, holme1, iyer1}, or betweenness centrality \cite{bellingeri1, norrenbrock1, almeira1, crucitti1, holme1, iyer1}, even though other types of attacks have been studied in the literature \cite{bellingeri1, iyer1, shao1, geng1}. 


The types of network mostly studied in the literature on the robustness of networks under attacks are the Erd\H{o}s-R\'enyi random graph \cite{albert1, bellingeri1, almeira1, crucitti1, holme1, shao1}, with exponential degree distribution, the Barabasi–Albert model \cite{albert1, bellingeri1, crucitti1, holme1}, with scale-free degree distribution, and the Watts–Strogatz model \cite{jalili1, holme1}, with intermediate features.

In a large body of literature the efficiency of the attack is measured in terms of the decrease of the size of the giant component \cite{ albert1, bellingeri1, norrenbrock1, almeira1, holme1, iyer1, shao1}, but less often more than one parameter is taken into account to measure network robustness \cite{norrenbrock1}. Also, in most works the effect of the attack is studied by comparing the damaged network with the original network \cite{albert1, bellingeri1, jalili1, norrenbrock1, almeira1, crucitti1, holme1, iyer1, shao1} so that the knowledge of the latter is necessary in order to establish whether an attack has been performed.  

In this short note we explore a simple machine-learning procedure to establish whether and how a given network has been attacked, without requiring the knowledge of the structure of the network before the attack. In short, we characterize each graph by a list of four normalized metrics: the ratio between the average and the maximum degree, the global clustering coefficient, the ratio between the average path length and the diameter, and the assortativity. While this list is not exhaustive, we show that it has statistically significant differences between intact and damaged networks, which suggests that this reduced list can indeed be used to detect damaged networks.

Focusing then on three basic random graphs, Erd\H{o}s-R\'enyi, Barabasi-Albert and Watts-Strogatz, we train two popular classification algorithms, $k$-Nearest-Neighbor and Random Forest, to recognize whether a given network has been attacked as well as the type of attack. We test our procedure on both artificial and real networks, first performing either targeted attacks, i.e. maximum-degree or maximum-betweenness node deletion, or random failures, and then applying our classification scheme to the resulting network. 

Even though the training set used in this paper is quite limited, our procedure is surprisingly successful in identifying the network type and distinguishing between random failures or targeted attacks, and could therefore provide a basis for more sophisticated approaches to the diagnosis and detection of damaged networks.

\section{The models}
In this work we train the classification procedure using as benchmarks the Erd\H{o}s-R\'enyi, Barabasi-Albert and Watts-Strogatz random graphs, and measure the attack efficacy using normalized versions, described below, of four basic graph metrics: the maximum degree $d_\text{max}$, the global clustering coefficient $C$, the average path length $\ell$ and the assortativity $r$. The generation of the graphs and the calculation of their metrics was performed using the Python package NetworkX \cite{hagberg2008exploring}. We describe below the basis tools used in this work.

The Erd\H{o}s-R\'enyi (ER) random graph $G(n,p)$ with $n$ nodes and link probability $p$ has binomial degree distribution, which for large $n$ and $np=\lambda$ constant reduces to the Poisson distribution with expectation $\lambda$. Here we choose $p=0.015,0.020,0.025,0.030$ and, for this and all the following graphs, $n$ of order $10^3$.

The Barabasi-Albert (BA) random graph with $n$ nodes and parameter $m$, denoted here by $B(n,m)$, is obtained by means of an iterative procedure as follows: starting from a star graph with $m+1$ nodes, a new node is added to the network at each step and linked to $m$ existing nodes, with probability proportional to their degree. The procedure is then iterated until order $n$ is attained. We choose here $m=5,6,7,8$.

Finally, the Watts-Strogatz (WS) random graph $W(n,k,p)$ with $n$ nodes, initial degree $k$ and rewiring probability $p$, is the random graph obtained by rewiring with probability $p$ the edges of a regular circle graph with degree $k$ and order $n$. Here we choose $p=0.1,0.2$ and $k=4,6,10$.

We briefly recall below the definitions of the metrics used in this work. For a undirected graph $G$ with order $n$, we denote by $\deg(i)$ the degree of node $i=1,\dots,n$, so that the maximum and average degrees are $d_\text{max}=\max_{i=1,\dots,n}\deg(i)$ and $d_\text{ave}=\sum_{i=1}^{n}\deg(i)/n$. 

The betweenness of a node $i$ is defined as $\beta(i)=\sum_{h,k=1}^n\frac{P(h,k;i)}{P(h,k)}$ where, for $h,k\ne i$, $P(h,k;i)$ is the number of shortest paths joining $h$ and $k$ that contain $i$, and $P(h,k)$ is the total number of shortest paths joining $h$ and $k$.

We write $C=\frac1n \sum_{i=1}^n\frac{N_i}{\deg(i)(\deg(i)-1)}$ for the global clustering coefficient, with $N_i$ the number of edges connecting the first neighbors of node $i$. 

The average path length is defined as $\ell=\frac1{n(n-1)} \sum_{j=1}^nd(i,j)$, with $d(i,j)$ the distance between nodes $i$ and $j$, and the diameter $\text{diam}(G)$ is the maximum distance between nodes in the graph. 

The assortativity, as introduced in \cite{newman}, is a measure of the correlation between the degrees of adjacent nodes, and relies on the notion of 'remaining degree', i.e. the number of edges exiting from a node connected to a random edge, the given edge being excluded. Letting $q(k)$ is the distribution of nodes with remaining degree $k$, $\sigma^2_q$ its variance, and $e(k,h)$ the joint distribution the remaining degrees of pair of nodes connected by a random edge, the assortativity is defined by $
r=\sum\limits_{k,h=1}^{n-1}\frac{kh(e(k,h)-q(k)q(h))}{\sigma^2_q}$, and takes values in $[-1,1]$.

\subsection{Normalized metrics} 
In order to compare graphs with different numbers of nodes, we use here normalized versions of the maximum degree and average path length, defined as
$$
\delta= d_\text{ave}/d_\text{max},\qquad
\lambda=\ell/\text{diam}(G).
$$
Notice that $\delta,\lambda\in[0,1]$, and that $\delta$ is inversely proportional to the maximum degree, so that it is expected to increase under attacks. Its usefulness in this context relies on the fact that targeted attacks affect the maximum degree more severely than the average degree. In the same spirit, since the diameter is more sensitive than the average path length under targeted attacks, we expect $\lambda$ to decrease in damaged networks, even though the average path length does increase under attack.

Henceforth, each graph in this paper will be identified by the following list of graph metrics:
\begin{equation}
(\delta,C,\lambda,r),
\label{list_of_metrics}
\end{equation}
i.e. the normalized reciprocal maximum degree, the global clustering coefficient, the normalized average path length, and the assortativity.


Since our goal is to compare networks with different orders and structural parameters, it may be useful to understand how the normalized metrics depend on these parameters. Analytical expressions are indeed available for ER and BA, less so for WS. These are summarized in Table \ref{tabella_formule} and briefly discussed below. 

\begin{table}[ht]
\begin{center}
    \begin{tabular}
{  |c|c|c|c|}%
 \hline%
  & $G(n,p)$ & $B(n,m)$ & $W(n,k,p)$ \\
 \hline
 $\delta$ & $\frac{pn}{pn + \sqrt{2p(1-p)n\log n}\left(
 1-\frac{\log\log n}{4\log n}-\frac{\log(2\pi^{1/2})}{2\log n}+\frac{y}{2\log n}
\right)}$ & $\frac{2mc}{\sqrt{n}}$ & $-$ \\
\hline
$C$ & $p$ & $\frac{\log^2n}{n} \text{ or } n^{-0.75}$ & $\frac{3(k-1)}{2(2k-1)}(1-p)^3$ \\
\hline
$\lambda$ & $\frac{\frac{\log n-\gamma}{\log pn}+\frac12}{
\frac{\log n}{\log pn}+2\frac{(10c/(\sqrt{c}-1)^2+1)}{c-\log 2c}\frac{\log n}{np}+1}$ & $\frac1{\text{diam}(G)}\left(\frac{\log n-\log (m/2)-1-\gamma'}{\log\log n+\log(m/2)}+ \frac{3}{2}\right)$ & $-$ \\
\hline
$r$ & $0$ & $-c_1\frac{\log^2n}{\sqrt{n}}\le r \le  c_2 n^{-\frac18}$ & $ - $ \\
\hline
\end{tabular}
\end{center}
\caption{Asymptotic estimates for the normalized metrics used in this work, see the main text for details. Missing entries correspond to lacking analytical estimates to our knowledge.}
\label{tabella_formule}
\end{table}
The asymptotic estimate for the average maximum degree of the ER random graph $G(n,p)$ follows from the expression for the max-degree distribution due to \cite{bollobas,riordan}, while those for the diameter and the average path length are obtained in \cite{fronczak} and \cite{chung}; in the formulae in the first column of Table \ref{tabella_formule} $y=0.577$ is the mean of the distribution $f(x)=e^{-x-e^{-x}}$, $\gamma$ is the Euler constant and $c\sim pn$ is a  constant. We are not aware of analytical expressions for the assortativity which, according to numerical simulations, is slightly negative for very small $p$ but tends to zero as $p$ increases \cite{vanmieghem}.

Regarding the scale-free network $B(n,m)$, the classical estimate for the average maximum degree can be obtained directly from the power-law degree distribution, with now $\gamma'\sim 3$ the power-law exponent and $c$ a suitable constant \cite{barabasibook}. The approximations for the clustering coefficient are due to \cite{bollobas} for a non-equivalent definition $\tilde C$ of the clustering coefficient, involving triangles instead of links between first neighbors, while the second estimate has been suggested by numerical simulations. The formula for the average path length has been derived in \cite{fronczak}, but we are not aware of analytical expressions for the diameter. For the assortativity of the BA graph lower and upper bounds have been estimated in \cite{shergin}.

As to the Watts-Strogatz random network $W(n,k,p)$, less is known regarding the metrics we are interested in. The degree distribution admits an analytical representation that shows an exponential decay for large degrees with average the ring-lattice degree $k$ \cite{barrat}. Therefore, the behavior of the normalized inverse maximum degree is expected to be similar to the ER graph, a suggestion confirmed by our numerical calculations. No analytical formula for the clustering coefficient is available to our knowledge, but there does exist an estimate for a modified version of this metric, in which the average number of links among the neighbors of a node is divided by the maximum number of possible links \cite{barrat}, and it can be shown that it matches the values of the true clustering coefficient with reasonable accuracy.

Even though the asymptotic expressions in Table \ref{tabella_formule} depend on unknown constants, they can be used to show that, in the range of networks orders we are interested into, the normalized metrics are essentially independent of $n$, so that it is reasonable that these can be applied to classify graphs with different number of nodes.

\subsection{Attacks}
We perform three types of attacks, consisting in the removal of a given fraction of nodes (here 1, 5 and 10\%) according to the following criteria: random failure, in which the nodes to be removed are drawn from a uniform distribution, and targeted attacks, in which the deleted nodes are those with decreasing degree or betweenness, starting with those with maximum degree or betweenness. It is known that targeted attacks are more efficient when the metrics are recomputed after each deletion \cite{bellingeri1, almeira1, holme1, iyer1}: here, for simplicity, we restrict to synchronous attacks, in which the nodes are removed simultaneously according to their centrality.

The procedure is as follows: we first generate intact random graphs with $n=500$, $800$ and $1000$ nodes. Then, the assigned fraction of nodes is removed, so that the attacked graphs have $495$, $475$, $450$ or $792$, $760$, $720$ or $990$, $950$, $900$ nodes. If the attack disconnects the network, this is discarded, so that only connected graphs are retained. Subsequently, intact ER, BA and WS random graphs are generated with these new numbers of nodes. These sets of intact and attacked networks are those that serve as benchmarks, as explained below. 

\section{The classification procedure}

\subsection{Hotelling test}
As a first sanity check to determine whether the list \eqref{list_of_metrics} is indeed suitable to characterize the effect of the attacks, we have performed a Hotelling test (Python package developed by F. Dion) comparing intact and attacked networks with the same parameters and numbers of nodes. The Hotelling test is a multivariate generalization of the Student $t$-test: the normalized reciprocal maximum degree, the average clustering coefficient, the normalized average path length, and the assortativity of two samples of 100 instances of intact and attacked networks of the same type (i.e. for instance, 100 intact and 100 attacked ER graphs with the same number of nodes and the same linking probability) are compared and the $p$-value is computed, the null hypothesis being that the samples are drawn from the same distribution. The results are summarized in Table \ref{hotelling}.

\subsection{Classification models}

Our goal here is to devise a procedure that allows to determine whether and how a given network has been attacked. To do this we use supervised learning algorithms, a subcategory of machine learning in which the dataset are labeled. The goal of the algorithms is to correctly establish the class of an unknown network, i.e. to determine its label.


In our work the data are lists \eqref{list_of_metrics} of metrics of a specific network, and the labels contain the following information:
\begin{itemize}
 
 \item $L_1$ characterizes the type of graph, i.e. ER, BA or WS.
 

 \item $L_2$ characterizes the type of attack and can assume four possible values: intact (i), random failure (r), maximum degree (d), or maximum betweenness (b).
 
 \item $L_3$ corresponds to the attack intensity, i.e. the fraction of removed nodes (1, 5 or 10\%).
\end{itemize}

Basically, in supervised learning input data are sequentially entered into a model and, at each stage, if the predictions are correct and the labels provided by the algorithm correspond to the true labels of the data, the procedure is iterated with new input data; otherwise, the parameters of the model are adjusted and the procedure is repeated until optimization.

Each model contains hyperparameters that cannot be modified in this way, so that in order to choose the optimal classification scheme the dataset is divided into three parts: the training, the validation and the test sets, as sketched below. 

The training set is used to train the model using given values of the hyperparameters. The model is then applied to the validation set, i.e. data that are unknown to the algorithm but whose true labels are known, in order to estimate the generalization error, which is the error that the model makes on data that has never seen. This error is evaluated using the accuracy, i.e. the fraction of correct predictions relative to the total number of tested samples. Subsequently, the algorithm is re-trained with different values of the hyperparameters, resulting in another generalization error. The optimal set of hyperparameters is usually found using a telescopic search technique: the hyperparameters are initially chosen in a wide range, and after completing the procedure with these values, this is iterated in a narrower range containing the hyperparameters that give the best result. The process is then iterated to the finest scale, resulting in a good estimate of the optimal hyperparameters, i.e. those that generalize better. 

The validation set is also used to find the optimal algorithm. Repeating the above procedure for different algorithms, the one with the smallest generalization error is chosen. In our work we have tested two algorithms: $k$-Nearest Neighbors and Random Forest.

Once the best algorithm is found, this is applied one last time on the training set using the best hyperparameters, after which the model is tested on the last portion of the dataset, the test set. These are data that have never been used during the procedure, and are used to estimate the final generalization error.

In our case the dataset split was $100-20-20$, i.e. for the training set 100 networks of each type were created - for example, 100 intact ER networks with $n$ = $495$ and $p$ = $0.015$, other 100 intact ERs with $n$ = $495$ and $p$ = $0.020$ and so on - 20 for the validation set and 20 for the test set.

\subsubsection{The $k$-Nearest Neighbors classifier}

As a first attempt, we have used the $k$-Nearest Neighbors classification scheme. This algorithm does not require a training phase: to each network in the training set is associated the list of normalized metrics, so that the dataset is a set of points in 4-dimensional space. The correct labels of these data points are known. During the validation phase the metrics of an unknown network are computed, and the labels of the closest $k$ data points are used to classify it. This can happen in two ways: either by a majority rule, i.e. the label most common in the subset of the closest data points is assigned to the unknown network, or the $k$ labels are weighted with an importance proportional to the reciprocal of the distance from the unknown data. The hyperparameters of this model are the number $k$ of nearest neighbors and the weights relative to the data, whether uniform or weighted by the reciprocal of the distance. The validation set is used to optimize the parameter $k$ - using telescopic search - and the type of distance, and the accuracy, which is the fraction of points that have been correctly classified, is used to estimate the generalization.


\subsubsection{The Random Forest classifier}

The Random Forest classifier is probably one of the most used classification algorithms and is based on the structure of decision trees. In a decision tree each node tests a condition on the input data, separating the tree into two or more branches. For example, for continuous variables, the condition could be whether some component of the data is above or below a certain threshold. With each ramification certain classes become more probable and others less likely; the final decision of the class to which the unknown data belong is taken at the terminal nodes, called leaf nodes.

\begin{figure}[ht]
	\begin{center}
	\includegraphics[width=1.1\textwidth]{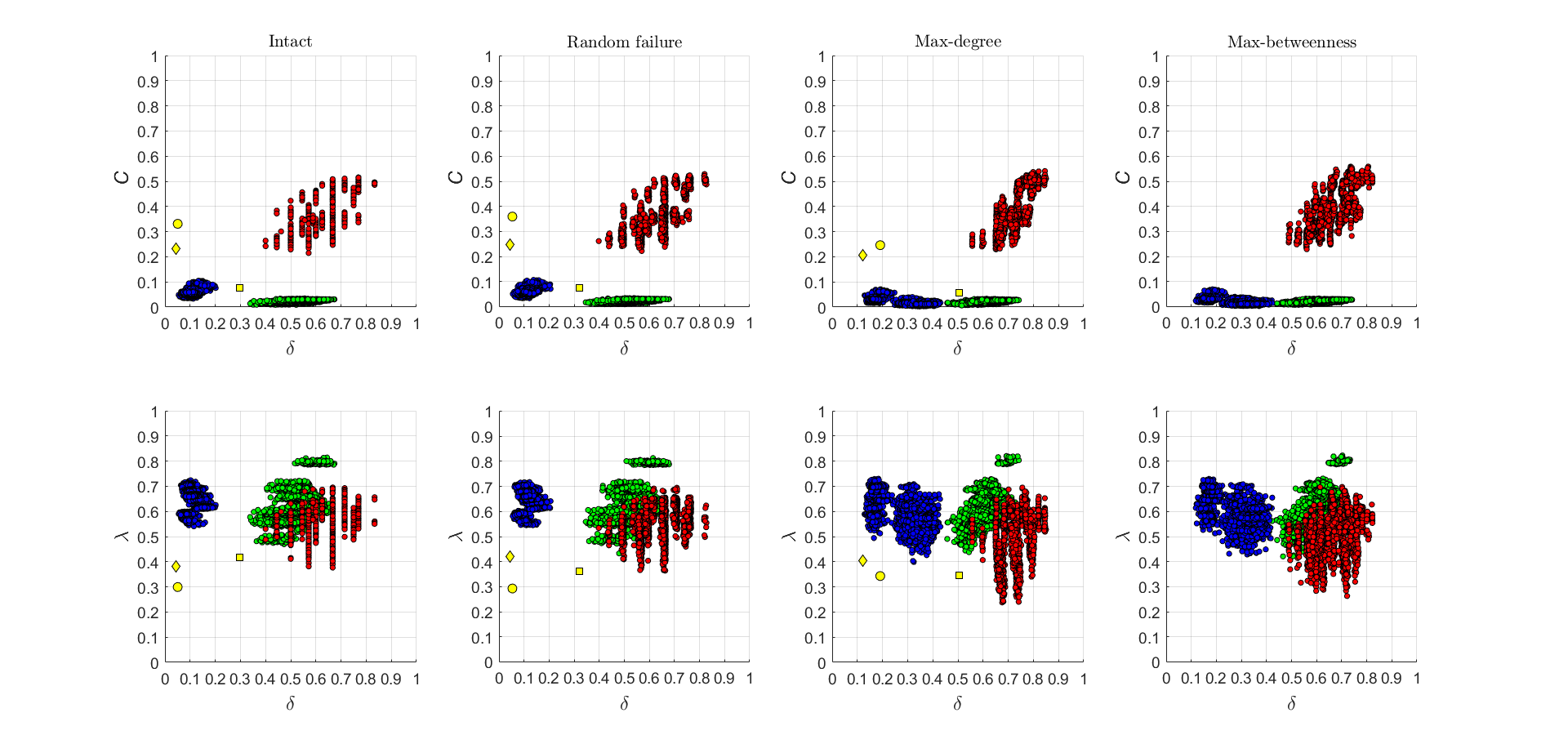}
	\end{center}
	\caption{Data points used to train the $k$-Nearest Neighbors classifier. Each data point corresponds to a network, colored in green for ER, in blue for BA and in red for WS. Data points are represented in the $(\delta,C)$ and $(\delta, \lambda)$ planes because the assortativity is very low in all cases. Notice that $\delta$ increases and $\lambda$ decreases in attacked graphs relative to the intact ones.  Also, the yellow circle, square and diamond are the data points corresponding to the intact and damaged fb-pages-food, power-bcspwr09 and web-polblogs networks, respectively (cf. Section \ref{real_networks}).}  	
	\label{networks}
\end{figure}
To create the best tree we rely on the gain of information, i.e. the difference in entropy before and after a given subdivision; the greater the information gain, the better the division, as it gets closer to the correct class. However, there is no unique way to create the tree: starting by applying the division from a certain metric rather than another, different trees are obtained with, eventually, different classifications.

The Random Forest classifier is based on $N$ decision trees that are independent; each decision tree predicts a certain class and the final prediction of the model is given by the class selected by the majority of the $N$ trees.

The strength of this model lies in the independence of the various trees; in fact, the prediction of the whole is more accurate than that of the single ones. This could be because some trees will be better at predicting certain classes correctly but will be wrong with others, and other trees vice versa.

In this case the algorithm contains only one hyperparameter, namely the number $N$ of decision trees.


\section{Results}

\subsection{Hotelling test}

The results of the statistical tests are summarized in Table  \ref{hotelling}, which shows the outcomes of a Hotelling test and 4 Student tests comparing the metrics of intact graphs to the those for the same graphs damaged by either random failures or maximum-degree or maximum-betweenness attacks.


Each group of three rows corresponds to a statistical test; the first three rows correspond to a Hotelling multivariate test aimed at determining whether there is a significant difference between the multivariate means of the 4 metrics of the intact and damaged network. The remaining 4 groups of 3 rows show the results of a Student test comparing the means of the corresponding metric of the intact and attacked network.

In all cases shown here, attacked networks with $n=990,950$ and $900$ nodes, obtained by deleting a fraction of 1\%, 5\% and 10\% nodes from ER, BA and WS graphs with 1000 nodes, have been compared with intact networks with the same number of nodes.

\begin{table}[ht]
\begin{center}
\begin{tabular}{|c|c|c c c|c c c|c c c|}
 \cline{3-11}  
 \multicolumn{2}{c|}{} 
  &  \multicolumn{3}{|c|}{\tiny Erd\H{o}s-R\'enyi}  &  \multicolumn{3}{c|}{\tiny Barabasi-Albert} & \multicolumn{3}{c|}{\tiny Watts-Strogatz}   \\
 \cline{3-11}  
 \multicolumn{2}{c|}{}  & \tiny $r$ & \tiny $d$ & \tiny $b$ & \tiny $r$ & \tiny $d$ & \tiny $b$ & \tiny $r$ & \tiny $d$ & \tiny $b$ \\ \thickhline
  \cline{1-11}  \multicolumn{1}{|c|}{}&\tiny   1\% &\tiny   1 &\tiny  4 &\tiny   4 &\tiny   0 &\tiny  4 &\tiny   4 &\tiny   3 &\tiny  6 &\tiny   6\\
 \cline{2-11}   $(\delta,C,\lambda,r)$   &\tiny  5\% &\tiny  2 &\tiny  4 &\tiny   4 &\tiny   0 &\tiny  4 &\tiny   4 &\tiny   5 &\tiny  6 &\tiny   6\\
    \cline{2-11}                  &\tiny  10\% &\tiny  0 &\tiny  4 &\tiny  4 &\tiny   0 &\tiny  4 &\tiny   4 &\tiny   6 &\tiny  6 &\tiny   6\\\thickhline
  \cline{1-11}  \multicolumn{1}{|c|}{}&\tiny   1\% &\tiny  1 &\tiny  4 &\tiny  4 &\tiny   0 &\tiny  4 &\tiny   4 &\tiny   0 &\tiny  6 &\tiny   6\\
 \cline{2-11}  $\delta$   &\tiny  5\% &\tiny  2 &\tiny  4 &\tiny   4&\tiny   0 &\tiny  4 &\tiny   4 &\tiny   0 &\tiny  6 &\tiny   6\\
    \cline{2-11}                  &\tiny  10\% &\tiny  0 &\tiny  4 &\tiny  4 &\tiny   0 &\tiny  4 &\tiny   4 &\tiny   0 &\tiny  6 &\tiny   6\\\thickhline
      \cline{1-11}  \multicolumn{1}{|c|}{}&\tiny   1\% &\tiny  0 &\tiny  4 &\tiny  4 &\tiny   0 &\tiny  4 &\tiny   4&\tiny   2 &\tiny  6 &\tiny   6\\
 \cline{2-11}    $C$   &\tiny  5\% &\tiny  0 &\tiny  4 &\tiny   4&\tiny   0 &\tiny  4 &\tiny   4&\tiny   5 &\tiny  6 &\tiny   6\\
    \cline{2-11}                  &\tiny  10\% &\tiny  0 &\tiny  4 &\tiny  4 &\tiny   0 &\tiny  4 &\tiny   4&\tiny   6 &\tiny  6 &\tiny   6\\\thickhline
      \cline{1-11}  \multicolumn{1}{|c|}{}&\tiny   1\% &\tiny  1 &\tiny  3 &\tiny  3 &\tiny   0 &\tiny  4 &\tiny   4 &\tiny   2 &\tiny  5 &\tiny   6\\
 \cline{2-11}   $\lambda$   &\tiny  5\% &\tiny  0 &\tiny  4 &\tiny   4 &\tiny   0 &\tiny  4 &\tiny   4 &\tiny   3 &\tiny  6 &\tiny   6\\
    \cline{2-11}                  &\tiny  10\% &\tiny  0 &\tiny  4 &\tiny  4 &\tiny   0 &\tiny  4 &\tiny   4 &\tiny   6 &\tiny  6 &\tiny   6\\\thickhline
      \cline{1-11}  \multicolumn{1}{|c|}{}&\tiny   1\% &\tiny  0 &\tiny  0 &\tiny  2 &\tiny   0 &\tiny  4 &\tiny   4 &\tiny   3 &\tiny  6 &\tiny   5\\
 \cline{2-11}  $r$   &\tiny  5\% &\tiny  1 &\tiny  0 &\tiny   4&\tiny   0 &\tiny  4 &\tiny   4&\tiny   5 &\tiny  6 &\tiny   6\\
    \cline{2-11}                  &\tiny  10\% &\tiny  0 &\tiny  0 &\tiny  4 &\tiny   0 &\tiny  4 &\tiny   4 &\tiny   5 &\tiny  6 &\tiny   6\\\thickhline
   \cline{1-11}  
\end{tabular}

\end{center}
\caption{Results of a Hotelling test and 4 Student tests comparing the metrics of intact graphs with the corresponding metrics for the same graphs damaged by either random failures ($r$), max-degree ($d$) or max-betweenness ($b$) attacks. Each entry in the table corresponds to a graph, an attack type $(r$, $d$, $b)$, intensity (1, 5, 10\%) and statistical test. For each entry, tests have been performed on 100 graphs generated using each of 4 constitutive parameters (the linking probabilities $p$ for ER and the numbers of attachment sites $m$ for BA) or 6 parameters (the rewiring probabilities $p$ and the initial degrees $k$ for WS). The number in each cell is the number of significant tests ($p$-value 0.05), i.e. those in which the attacked network shows a significant difference in mean with the intact graph.}
\label{hotelling}
\end{table}

We have performed tests for all values of the relevant parameters: 4 values $p=0.015,0.020,0.025,0.030$ for the Erd\H{o}s–R\'enyi graph $G(n,p)$, 4 values $m=5,6,7,8$ for the Barabasi-Albert graph $B(n,m)$, and 6 pairs of values $p= 0.1,0.2$ and $k=4,6,10$ for the Watts-Strogatz graph $W(n,k,p)$. The number in each cell is the number of attacked graphs whose corresponding metric (or list of metrics) shows a significant difference in mean with the intact graph ($p$-value less than 0.05). Therefore, for the ER, BA and WS graphs the numbers in each cell ranges from 0 to 4 or 6, respectively, corresponding to the values of $p$ (ER), $m$ (BA) or the pair $(p,k)$ (WS). 




The results in Table \ref{hotelling} show that, for both ER and BA graphs, the list of 4 metrics used in this work does show a significant difference between the intact graphs and those which have undergone an attack targeted at max-degree or max-betweenness nodes. On the other hand, the list is not generally able to tell between random failures and intact networks. 


The results in Table \ref{hotelling} for Watts-Strogatz, on the other hand, show a significant difference, in both the complete list and each individual metric, excluding the normalized degree; this applies to all types of attacks, including random failures, especially for the most intense attacks. This seems to indicate a substantial fragility of WS relative to the other graphs. 

In any event, the statistical tests confirm that the list of metrics used here are indeed capable to distinguish between intact and attacked networks. Table \ref{hotelling} corresponds to networks with 1000 nodes before the attack; we have performed the same test for $n=800$ and $n=500$ with similar results.

\subsection{Classification models}

Since the $k$-Nearest Neighbors algorithm yields a lower accuracy than Random Forest, we only discuss here the performance of the latter, as evaluated both by the accuracy of the results, i.e. the fraction of correct predictions in the test set, and by the confusion matrix, which is a more refined performance evaluation tool than the accuracy. The confusion matrix is a matrix whose generic entry $C_{ij}$ is the number of networks that we know to belong to class $i$ and which the model has classified as belonging to $j$. The two quantities are related, since the accuracy is the sum of the elements on the main diagonal of the confusion matrix divided by the total number of the elements. In Tables \ref{conf1} and \ref{conf2}, this relationship is not evident as the values have been normalized to make the matrix stochastic by rows. To find the relationship between accuracy and confusion matrix, the accuracy of each row must be multiplied by the ratio between the number of data in the row divided by the number of total data.

In this work, we have applied the classification procedure at the finest scale, but we group below some classes together in order to make the results more understandable and generalisable.

\begin{itemize}
    \item The fine classification procedure allows to distinguish the type of graph, the type of attack and its intensity, so that each network is classified according to the values of $L_1=ER,BA,WS$, $L_2=i,r,d,b$ and $L_3=1\%,5\%,10\%$. At this level of classification, the accuracy of the algorithm is 0.6193.



\item By neglecting the information regarding the attack intensity, and using the labels $L_1=ER,BA,WS$ and $L_2=i,r,d,b$ only, we obtain a second, coarser classification. For example, if two networks were previously classified as $ER/i/1\%$ and $ER/i/5\%$, they would now be in the same class $ER/i$.
Now the accuracy is 0.7178 and the related confusion matrix is reported in Table \ref{conf1}.

\item 

Finally, by grouping together intact and randomly-attacked graphs in a single class, and targeted attacks (degree and betweenness) in another, we obtain the coarsest classification of the data. Here the attack intensity is again neglected, so that the only label is $L_{12}$ with values $ER/(i,r)$, $ER/(d,b)$, $BA/(i,r)$, $BA/(d,b)$, $WS/(i,r)$, $WS/(d,b)$.
For this classification the accuracy rises to 0.9713, which yields the confusion matrix in Table \ref{conf2}.
\end{itemize}

\landscape
\begin{table}[h]
\begin{tabular}{c|c|c|c|c|c|c|c|c|c|c|c|c|}
\cline{2-13}
\textbf{}                           & \textbf{BA/i} & \textbf{ER/i} & \textbf{WS/i} & \textbf{BA/r} & \textbf{ER/r} & \textbf{WS/r} & \textbf{BA/d} & \textbf{ER/d} & \textbf{WS/d} & \textbf{BA/b} & \textbf{ER/b} & \textbf{WS/b} \\ \hline
\multicolumn{1}{|c|}{\textbf{BA/i}} & 0.5222        & 0.0           & 0.0           & 0.4778        & 0.0           & 0.0           & 0.0           & 0.0           & 0.0           & 0.0           & 0.0           & 0.0           \\ \hline
\multicolumn{1}{|c|}{\textbf{ER/i}} & 0.0           & 0.4986        & 0.0           & 0.0           & 0.45          & 0.0           & 0.0           & 0.025         & 0.0           & 0.0014        & 0.025         & 0.0           \\ \hline
\multicolumn{1}{|c|}{\textbf{WS/i}} & 0.0           & 0.0           & 0.988         & 0.0           & 0.0           & 0.0093        & 0.0           & 0.0           & 0.0           & 0.0           & 0.0           & 0.0028        \\ \hline
\multicolumn{1}{|c|}{\textbf{BA/r}} & 0.5042        & 0.0           & 0.0           & 0.4944        & 0.0           & 0.0           & 0.0014        & 0.0           & 0.0           & 0.0           & 0.0           & 0.0           \\ \hline
\multicolumn{1}{|c|}{\textbf{ER/r}} & 0.0           & 0.5111        & 0.0           & 0.0           & 0.4417        & 0.0           & 0.0           & 0.0125        & 0.0           & 0.0014        & 0.0333        & 0.0           \\ \hline
\multicolumn{1}{|c|}{\textbf{WS/r}} & 0.0           & 0.0           & 0.0046        & 0.0           & 0.0           & 0.9167        & 0.0           & 0.0           & 0.0241        & 0.0           & 0.0           & 0.0546        \\ \hline
\multicolumn{1}{|c|}{\textbf{BA/d}} & 0.0           & 0.0           & 0.0           & 0.0           & 0.0014        & 0.0           & 0.6167        & 0.0           & 0.0           & 0.3819        & 0.0           & 0.0           \\ \hline
\multicolumn{1}{|c|}{\textbf{ER/d}} & 0.0           & 0.0097        & 0.0           & 0.0           & 0.0125        & 0.0           & 0.0           & 0.6264        & 0.0           & 0.0           & 0.3514        & 0.0           \\ \hline
\multicolumn{1}{|c|}{\textbf{WS/d}} & 0.0           & 0.0           & 0.0           & 0.0           & 0.0           & 0.0093        & 0.0           & 0.0           & 0.9694        & 0.0           & 0.0           & 0.0213        \\ \hline
\multicolumn{1}{|c|}{\textbf{BA/b}} & 0.0           & 0.0           & 0.0           & 0.0           & 0.0           & 0.0           & 0.4028        & 0.0           & 0.0           & 0.5972        & 0.0           & 0.0           \\ \hline
\multicolumn{1}{|c|}{\textbf{ER/b}} & 0.0           & 0.0264        & 0.0           & 0.0           & 0.025         & 0.0           & 0.0014        & 0.3708        & 0.0           & 0.0           & 0.5764        & 0.0           \\ \hline
\multicolumn{1}{|c|}{\textbf{WS/b}} & 0.0           & 0.0           & 0.0019        & 0.0           & 0.0           & 0.0574        & 0.0           & 0.0           & 0.0315        & 0.0           & 0.0           & 0.9093        \\ \hline
\end{tabular}
\caption{The confusion matrix relative to the classification in terms of $L_1$ and $L_2$. All values have been normalized so that the sum over each row is 1, so that the elements $C_{ii}$ indicate the accuracy in classifying the specific class $i$ and the elements $C_{ij}$, with $i\neq{j}$, the errors in classifying that class.}
\label{conf1}
\end{table}
\endlandscape

\begin{table}[h]
\begin{tabular}{c|c|c|c|c|c|c|}
\cline{2-7}
\textbf{}                             & \textbf{BA/i,r} & \textbf{ER/i,r} & \textbf{WS/i,r} & \textbf{BA/d,b} & \textbf{ER/d,b} & \textbf{WS/d,b} \\ \hline
\multicolumn{1}{|c|}{\textbf{BA/i,r}} & 0.9993          & 0.0             & 0.0             & 0.0007          & 0.0             & 0.0             \\ \hline
\multicolumn{1}{|c|}{\textbf{ER/i,r}} & 0.0             & 0.9507          & 0.0             & 0.0014          & 0.0479          & 0.0             \\ \hline
\multicolumn{1}{|c|}{\textbf{WS/i,r}} & 0.0             & 0.0             & 0.9593          & 0.0             & 0.0             & 0.0407          \\ \hline
\multicolumn{1}{|c|}{\textbf{BA/d,b}} & 0.0             & 0.0007          & 0.0             & 0.9993          & 0.0             & 0.0             \\ \hline
\multicolumn{1}{|c|}{\textbf{ER/d,b}} & 0.0             & 0.0368          & 0.0             & 0.0007          & 0.9625          & 0.0             \\ \hline
\multicolumn{1}{|c|}{\textbf{WS/d,b}} & 0.0             & 0.0             & 0.0343          & 0.0             & 0.0             & 0.9657          \\ \hline
\end{tabular}
\caption{ The confusion matrix relative to the coarsest label $L_{12}$. Also in this case the values have been normalized (see Table \ref{conf1}).}
\label{conf2}
\end{table}

The remarkable increase in accuracy from Tables \ref{conf1} to \ref{conf2} demonstrates that the main error of the algorithm is in distinguishing intact networks from those that have undergone random failures and between the two targeted attacks. Instead, the errors related to the recognition of the type of graph or the distinction between targeted and intact-random attacks are minimal.

It may also be important to remark that, for the Watts-Strogatz graph, the accuracy is already quite high in the fine classification procedure, which is able to distinguish between the intact network and random failures, as well as between max-degree and max-betweenness attacks.

\subsection{Real networks}\label{real_networks}

Although the classification algorithm was trained on networks generated by theoretical models only, we have applied it to some real networks as well, based on the (hazardous) assumption that at least one of the three models behave similarly to the real network.

We chose three networks \cite{net_rep} from contexts as dissimilar as possible but where a random failure and a targeted attack made sense. The first, fb-pages-food, is a social network made up of Facebook pages relating to different categories, in which the nodes are the pages and the links are mutual likes among them. The second, power-bcspwr09, is a power network, while the third, web-polblogs, is a web graph.

We have applied to these networks both random and maximum-degree attacks deleting 10\% of the nodes; in terms of the coarsest classification, with label $L_{12}$, the two networks obtained from the attack of the social network have been correctly classified as $WS/i,r$ and $WS/d,b$. Similarly, the networks obtained from the power network were classified as $ER/i,r$ and $ER/d,b$ and, finally, those deriving from the web graph as $WS/i,r$ and $WS/d,b$. 

Even though the procedure always associates the same graph label $L_1$ to each pair of attacked networks, in these cases the interesting outcome does not lie in the type of graph that has been predicted, but in the correct assessment of the type of attack that the network has undergone.

\section{Discussion}

Given that a network is a structure that describes the relations among functionally interconnected entities,
it seems reasonable that the main purpose of a targeted attack aimed at disrupting the function of a network is to decrease its connectivity. From this point of view, it is natural to quantify the damage in terms of the size of the giant component, as done in the first studies on the robustness and the resilience of networks.      

However, a targeted attack affects the topology of a network in many subtle ways, which are difficult to quantify given the vast structural and functional diversity of the networks encountered in applications. For instance, the small world property is not simply related to the connectivity alone, nor is the assortativity, a measure on the tendency of nodes to link to other nodes with similar degree. Further, the effect of an attack on a network depends on its function, which is the feature to which a targeted attack is directed.

On the other hand, many targeted attacks follow a simple logic such as, for instance, the removal of high-connectivity nodes, and the question arises as to whether there are common features in the damage done by these to networks with very different structures. Given that many of the relevant topological features of massive network are summarized by the so-called graph-metrics, the hope is that the damage can be quantified in terms of these metrics. 


Hence, in this paper we explore an automated approach to the task of determining whether a given network has been attacked, based on two popular machine-learning algorithms. More precisely, we examine a simplified setup, in which the algorithm is trained using three popular random graphs, Erd\H{o}s-R\'enyi, Barabasi-Albert and Watts-Strogatz, and targeted attacks based on the removal of maximum-degree or maximum-betweenness nodes, as well as random failures. Each graph is characterized by 4 scalar quantities: $\delta$, a measure of connectivity given by the ratio between the average and maximum degree; $C$, the average clustering coefficient; $\lambda$, the ratio between the average path length and the diameter and $r$, the assortativity.  

Since targeted attacks here consist in the removal of nodes with maximal connectivity, which affect the maxima of the degree and path-length distributions more severely than the corresponding averages, the quantities $\delta$ and $\lambda$ are consistently larger and smaller (respectively) in damaged networks relative to intact ones. Further, it turns out that, for a wide range of the number of nodes of the network, they are essentially independent of that number. However, random failures, in which the nodes are deleted according to an uniform distribution, do not significantly change the above metrics.
 
The first main result of our work is that the Random Forest classifier gives highly satisfactory results for the classical random graphs, and is able to identify correctly whether a graph has been attacked by a targeted attack, while it is unable to distinguish between intact and randomly-attacked networks. This could be due to the fact that the random graphs used for the training have special, artificially enhanced features, and therefore it could be comparatively easy to assess whether an attack has significantly affected these features. 

However, applying the classification algorithm to three real networks, both subject to random failures and targeted attacks, shows that, somewhat surprisingly, this is able to understand correctly whether the graph has been attacked.

This result suggests that machine-learning approaches may indeed be successful in performing the complex task of diagnosing whether a network has been attacked. Our work here, though, should be viewed as a preliminary exploration of this avenue, since for a more accurate and general application to real networks it would be necessary not to restrict the training to the three models used here, but definitely on other networks similar to those to be studied.


\section{Acknowledgements} P.C. acknowledges the support of the research project  ‘Stochastic and statistical models and methods for the applications’ (University of Torino, SACL-RILO-18-01).

\bibliography{bibliografia2}

\end{document}